\documentstyle[12pt]{article}
\newlength{\extraspace}
\newlength{\extraspaces}
\newcommand{\be}{\begin{equation}
\addtolength{\abovedisplayskip}{\extraspaces}
\addtolength{\belowdisplayskip}{\extraspaces}
\addtolength{\abovedisplayshortskip}{\extraspace}
\addtolength{\belowdisplayshortskip}{\extraspace}}
\newcommand{\ee}{\end{equation}}

\newcommand{\ba}{\begin{eqnarray}
\addtolength{\abovedisplayskip}{\extraspaces}
\addtolength{\belowdisplayskip}{\extraspaces}
\addtolength{\abovedisplayshortskip}{\extraspace}
\addtolength{\belowdisplayshortskip}{\extraspace}}
\newcommand{\ea}{\end{eqnarray}}

\newcommand{\arctanh}{{\mathop{\rm arctanh}\nolimits}}

\newcommand{\figone}[2]{
\begin{figure}
\vspace{#1mm}
\begin{center}
\setlength{\unitlength}{1.1mm}
\raisebox{-20\unitlength}
{\mbox{\begin{picture}(70,65)(-35,-35)
\thicklines
\put(-30,0){\line(1,1){30}}
\put(0,30){\line(1,-1){30}}
\put(30,0){\line(-1,-1){30}}
\put(0,-30){\line(-1,1){30}}
\put(-15,-15){\line(1,1){30}}
\put(15,-15){\line(-1,1){30}}
\put(-14.7,14.7){\line(1,0){29.4}}
\put(-14.7,15.3){\line(1,0){29.4}}
\put(-14.7,-14.7){\line(1,0){29.4}}
\put(-14.7,-15.3){\line(1,0){29.4}}
\put(15,0){\makebox(0,0){I}}
\put(-15,0){\makebox(0,0){I$^\prime$}}
\put(0,10){\makebox(0,0){II}}
\put(0,-10){\makebox(0,0){II$^\prime$}}
\put(0,20){\makebox(0,0){IV}}
\put(0,-20){\makebox(0,0){IV$^\prime$}}
\end{picture}}}
\parbox{13cm}{\small #2}
\end{center}
\end{figure}}

\newcommand{\figtwo}[2]{
\begin{figure}
\vspace{#1mm}
\begin{center}
\setlength{\unitlength}{1.1mm}
\raisebox{-20\unitlength}
{\mbox{\begin{picture}(80,170)(-40,-140)
\thicklines
\put(-30,0){\line(1,1){35}}
\put(-5,35){\line(1,-1){35}}
\put(30,0){\line(-1,-1){60}}
\put(30,-60){\line(-1,1){60}}
\put(-30,-30){\line(1,1){60}}
\put(30,-30){\line(-1,1){60}}
\put(-30,30){\line(1,1){5}}
\put(30,30){\line(-1,1){5}}
\put(-30,-30){\line(1,-1){35}}
\put(30,-30){\line(-1,-1){35}}
\put(30,-60){\line(-1,-1){5}}
\put(-30,-60){\line(1,-1){5}}
\thinlines
\multiput(-14.5,15)(3,0){10}{\line(1,0){2}}
\multiput(-14.5,-15)(3,0){10}{\line(1,0){2}}
\multiput(-14.5,-45)(3,0){10}{\line(1,0){2}}
\put(15,30){\makebox(0,0){I}}
\put(-15,30){\makebox(0,0){I}}
\put(15,0){\makebox(0,0){I}}
\put(-15,0){\makebox(0,0){I}}
\put(15,-30){\makebox(0,0){I}}
\put(-15,-30){\makebox(0,0){I}}
\put(15,-60){\makebox(0,0){I}}
\put(-15,-60){\makebox(0,0){I}}
\put(0,10){\makebox(0,0){II}}
\put(0,-10){\makebox(0,0){II}}
\put(0,20){\makebox(0,0){II}}
\put(0,-20){\makebox(0,0){II}}
\put(0,-40){\makebox(0,0){II}}
\put(0,-50){\makebox(0,0){II}}
\put(0,-75){\makebox(0,0){(a)}}
\thicklines
\put(7.8,-89.7){\line(-1,-1){15.3}}
\put(-7.5,-105){\line(1,-1){15.3}}
\put(7.8,-89.7){\line(0,-1){30.6}}
\put(7.2,-90.3){\line(0,-1){29.4}}
\put(2,-105){\makebox(0,0){IV}}
\put(0,-130){\makebox(0,0){(b)}}
\end{picture}}}
\parbox{13cm}{\small #2}
\end{center}
\end{figure}}

\begin{document}
\addtolength{\baselineskip}{1.5mm}

\thispagestyle{empty}
\begin{flushright}
DAMTP R93/1\\
hep-th/9302037
\end{flushright}
\vspace{.6cm}

\begin{center}
{\LARGE{Non-Singularity of the Exact \\[2mm]
        Two-Dimensional String Black Hole}}\\[16mm]
{Malcolm J.~Perry~ and~ Edward Teo}\\[6mm]
{Department of Applied Mathematics and Theoretical Physics\\[1mm]
University of Cambridge\\[1mm]
Silver Street\\[1mm]
Cambridge CB3 9EW\\[1mm]
England}
\end{center}
\vspace{1.5cm}

\centerline{\bf Abstract}
\noindent
We study the global structure of the exact two-dimensional space-time
which emerges from string theory. Previous work has shown that in the
semi-classical limit, this is a black hole similar to the Schwarzschild
solution. However, we find that in the exact case, a new Euclidean region
appears ``between'' the singularity and black hole interior. However the
boundary between the Lorentzian and Euclidean regions is a coordinate
singularity, which turns out to be a surface of time reflection symmetry in
an extended space-time. Thus strings having fallen through the black hole
horizon would eventually emerge through another one into a new asymptotically
flat region. The maximally extended space-time consists of an infinite
number of universes connected by wormholes. There are no singularities
present in this geometry. We also calculate the mass and temperature
associated with the space-time.

\vfill
\vspace{1cm}
\begin{flushleft}
{Feburary 1993}
\end{flushleft}


\newpage

One of the great problems of classical general relativity is that it
is plagued by space-time singularities. For example, one of the singularity
theorems \cite{Penrose} asserts that if one has a trapped surface in an
asymptotically flat space-time with matter obeying the strong energy condition
(as classical matter traditionally does), and which satisfies Einstein's
equation, then the space-time is geodesically incomplete. A simple
corollary of this is that a black hole must be associated with a space-time
singularity.

There has been some hope that a satisfactory quantum theory of gravity
will resolve the difficulties associated with singularities. The only
known realistic candidate for such a theory at present is string theory.
Contrary to popular folklore, our understanding of string theory has seen
important advances in the past few years. Perhaps most notable amongst
these has been the realization that consistent string theories can be
constructed in target space dimensions much lower than the critical
dimension. A very successful example is the $c=1$ matrix model (for reviews
see \cite{Ginsp,Kleb}), which non-perturbatively describes strings
propagating in a $1+1$-dimensional background. Another major advance
has been the discovery of background metrics for these low-dimensional
target spaces. These classical solutions are in general curved space-times,
and indeed the prototype is a black hole solution in $1+1$ dimensions
\cite{Wadia,WittenI}.

The dynamics of strings in a curved space-time is governed by the world-sheet
conformal invariance of string theory, which in turn is imposed by the
vanishing of the $\beta$-functions for the target space massless fields
\cite{Callan}. In $1+1$ dimensions for the bosonic string, these are the
metric, dilaton and tachyon. However the $\beta$-function equations are
only known perturbatively in the inverse string tension $\alpha^\prime$,
and so conformal invariance can only be imposed order by order. The
$1+1$-dimensional black hole of Ref.~\cite{Wadia}\ was found by setting
the tachyon to zero and solving the lowest order $\beta$-functions, which
have the form of Einstein's equation coupled to a dilaton.

The resulting black hole solution is given by the metric and dilaton
\cite{WittenI}
\ba
\label{BHi}
{\rm d} s^2&=&-\tanh^2\lambda r\ {\rm d} t^2+{\rm d} r^2\ ,\nonumber\\
\phi&=&\phi_0+\ln\cosh^2\lambda r\ ;
\ea
or, in Kruskal-type coordinates, as
\ba
\label{BHii}
{\rm d} s^2&=&-{1\over\lambda^2}\ {{\rm d} u{\rm d} v\over1-uv}\ ,\nonumber\\
\phi&=&\phi_0+\ln(1-uv)\ ,
\ea
for some constant $\lambda$ which depends on the central charge of the string
world sheet. It is clearly seen to possess a causal structure similar to that
of the Schwarzschild black hole in four dimensions \cite{WittenI}: with an
event horizon at $uv=0$ and a curvature singularity at $uv=1$. The Penrose
diagram for the maximally extended black hole space-time is shown in Fig.~1.

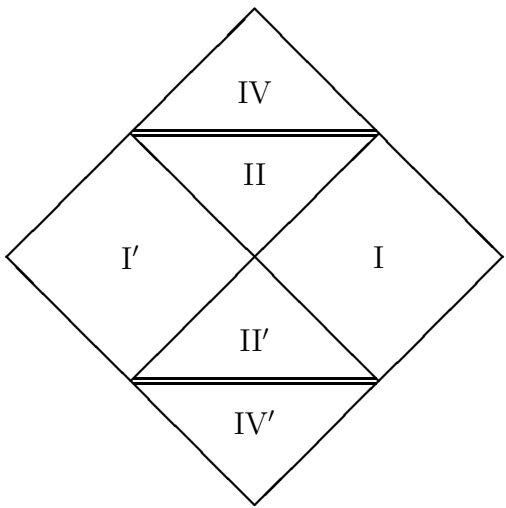
\figone{0}{Fig.~1. Penrose diagram for the two-dimensional black hole as
discussed by Witten. Regions I, I$^\prime$ are asymptotically flat space-times
exterior to the black hole and white hole horizons. Regions II, II$^\prime$
are inside the horizons, while IV, IV$^\prime$ are asymptotically flat regions
each containing a naked singularity. The double lines represent the curvature
singularities at $uv=1$.}

Witten has managed to find an exact conformal field theory description of
this black hole \cite{WittenI}, which ensures that conformal invariance is
obeyed non-perturbatively to all orders in $k$, the Kac--Moody level. This
description is in the form of a Wess--Zumino--Witten (WZW) model
\cite{WittenII}\ based on the non-compact group ${\rm G}={\rm SO}(2,1)$,
gauged by the non-compact group ${\rm H}={\rm SO}(1,1)$. The resulting
coset space construction G/H maps to a non-compact $1+1$-dimensional
gravitational background in which strings propagate. In the
$k\rightarrow\infty$
approximation, the above black hole solution is recovered \cite{WittenI}.
However, because of the relationship between $k$ and the central charge $c$
of the SO(2,1)/SO(1,1) model given by \cite{GKO}
\be
c={3k\over k-2}-1\ ,
\ee
$k$ takes the value $9/4$ for a bosonic string background since $c=26$
in order to cancel the contribution from the diffeomorphism ghosts.
Thus, $1/k$ is quite large, and corrections due to this should not be ignored.

The effective space-time background for general $k$ was first derived
by Dijkgraaf {\it et al.} \cite{Verlinde}, and is
\ba
\label{Exacti}
{\rm d} s^2&=&2(k-2)\left[-\beta(r)\,{\rm d}t^2+{\rm d}r^2\right]\ ,\nonumber\\
\phi&=&\phi_0+\hbox{$1\over2$}\ln\left(\sinh^22\lambda r/\beta(r)\right)\ ,
\ea
where $\beta(r)$ is given by
\be
\label{Exactii}
\beta(r)=\left(\coth^2\lambda r-{2\over k}\right)^{-1}.
\ee
It reduces to (\ref{BHi}) for $k\rightarrow\infty$ (up to an overall
conformal factor), and is believed to solve the $\beta$-function
equations exactly. This has been confirmed by explicit computation to
four loop level in Refs.~\cite{Tseytlin,Jack}.

A few things can be said about the geometry of this space-time for $k>2$,
which we will assume from now on. Apart from considerations of string
physics, it is believed that $k>2$ in order to have a unitary conformal
field theory. The coordinate patch (\ref{Exacti}) and (\ref{Exactii}) is
asymptotically flat and describes the geometry exterior to an horizon at
$r=0$. In fact there are two copies of this region, one for $r$ positive
and the other for $r$ negative, which are joined together at $r=0$ in a
way analogous to the Kruskal bridge of the Schwarzschild black hole.
This space-time corresponds to regions I and I$^\prime$ of Fig.~1. Also
note that the dilaton grows in strength asymptotically.

It is simple to see that the space-time given here is a black hole by a
straightforward analogy with the $k\rightarrow\infty$ limit. The black hole
horizon, at $r=0$, is as usual associated with a Hawking temperature. The
easiest way to find the temperature \cite{Gibbons} is to analytically continue
the metric (\ref{Exacti}) to its Euclidean region, and observe that the conical
singularity at $r=0$ is removed by identifying the Euclidean time coordinate
with period $2\pi/\lambda$. The inverse of the proper period at infinity
is then the Hawking temperature, and is given by $T_H=\lambda/(2\pi\sqrt{2k})$.

One can also estimate the mass of this black hole. Assuming that the target
space action is given by the Einstein action, and that the various higher-order
corrections are negligible as one goes towards $r\rightarrow\infty$, one
can calculate the ADM mass by following standard techniques. The result is
\be
M=2(k-2)\ \lambda e^{\phi_0}\left(1-\hbox{$2\over k$}\right)^{-3/2},
\ee
which agrees with the mass of the black hole in the $k\rightarrow\infty$
limit \cite{Perry}, if the conformal factor $2(k-2)$ is dropped.

We can now determine what the space-time represents physically.
Outside the horizon the space-time is static, and therefore it represents
a black hole in thermodynamic equilibrium with a cloud of tachyons at the
Hawking temperature $T_H$. Since the mass is finite, it seems that the
total energy of the radiation is finite. Thus the entire system is
gravitationally bound, rather like an isolated ``star''.

Within the context of gauged WZW models, there exists a duality transformation
\cite{Verlinde} which maps the above space-time to a new one again of the
form (\ref{Exacti}), but now with
\be
\label{Exactiii}
\beta(r)=\left(\tanh^2\lambda r-{2\over k}\right)^{-1}.
\ee
The conformal field theories associated to each of these two dual target spaces
are completely equivalent, but their geometries are very different. The
new space-time given by (\ref{Exactiii}) is also asymptotically flat,
but there is a curvature singularity at $r_c\equiv\lambda^{-1}\arctanh
\sqrt{2/k}$. Thus the region $r_c<r<\infty$, which we denote by IV and where
$\beta(r)>0$, is a space-time that is exposed to a naked singularity. There
is also a region behind the singularity $0\leq r<r_c$, where $\beta(r)<0$,
which we call III. It vanishes in the $k\rightarrow\infty$ limit, and so
has no analog in Fig.~1.

A new form of the metric which extends over the different regions of the
entire space-time has been found in Ref.~\cite{BarsI}. It is
\ba
\label{BHiii}
{\rm d} s^2&=&2(k-2)\left[-\beta(x)\,{\rm d}t^2
+{{\rm d} x^2\over 4(x^2-1)}\right]\ ,\nonumber\\
\beta(x)&=&\left({x+1\over x-1}-{2\over k}\right)^{-1}\ ,
\ea
with the associated dilaton
\be
\phi=\phi_0+\hbox{$1\over2$}\ln\left({x^2-1\over\beta(x)}\right)\ .
\ee
To recover the above metrics, we have to reparametrize $x$ in the various
regions of this extended geometry. By setting $x=\cosh2\lambda r\ge1$,
(\ref{BHiii}) reduces to the metric (\ref{Exacti}) with (\ref{Exactii}),
thus yielding the exterior region I. The black hole horizon is at $x=1$.
By contrast, the dual metric, consisting of regions III and IV, is
obtained via the redefinition $x=-\cosh2\lambda r\le-1$. The singularity
is at $x_c\equiv-(k+2)/(k-2)<-1$. Lastly, region II inside the horizon is
labeled by $-1<x<1$, and a suitable parametrization of it is $x=\cos2\lambda
r$.

Note that in addition to being singular at the horizon and singularity,
the metric (\ref{BHiii}) is singular at the point $x=-1$ between them.
It is a coordinate singularity as can be seen from the regularity of the
scalar curvature
\be
R={2k\over k-2}\ {(k-2)x+k-4\over \left[(k-2)x+k+2\right]^2}\ ,
\ee
at that point. Does it correspond to anything special?

{}From (\ref{BHiii}), observe that region I has signature $(-+)$, region II
has signature $(+-)$, region III has signature $(++)$, while region IV has
signature $(-+)$. Thus $x=-1$ appears to mark the boundary between two
regions of different space-time signature. But what does it mean to have
a region of Euclidean signature between the horizon and singularity of a
Lorentzian black hole? Specifically, what happens to a particle\footnote{In
a single spatial dimension, a string has no transverse oscillations, and so
its motion is completely specified by its center of mass coordinates
\cite{Martinec}. Thus strings resemble point particles in two dimensions.}
when it falls into the black hole and reaches $x=-1$? By studying geodesic
motion in this geometry, this boundary turns out to be at finite proper
distance from any point at finite $x$. However, nothing can enter the
Euclidean region III as the signature of the space-time is fixed to be
Lorentzian.

There is no real physical meaning attached to this ``boundary'', which
follows by finding coordinates which are non-singular there, e.g.,
\be
x=-1+2\rho^2\ .
\ee
Then the metric (\ref{BHiii}) takes the form
\be
{\rm d} s^2=2(k-2)\left[{k(1-\rho^2)\over2+\rho^2(k-2)}\,{\rm d}t^2
-{{\rm d}\rho^2\over1-\rho^2}\right]\ .
\ee
It is flat and non-singular at $\rho\simeq0$. Thus an infalling particle has
radial coordinate $\rho$ running smoothly through $\rho=0$, and it sees
nothing special at that point. However, there are coordinate singularities
at $\rho=\pm1$, corresponding to two copies of the horizon $x=1$.

To summarize, we have found that the black hole space-time is disconnected
from the singularity by a region of Euclidean signature. A particle falling
into the black hole sees no singularity, but emerges into another
asymptotically
flat black hole space-time. Two copies of region II are glued together at
$x=-1$ to form a wormhole bridging two asymptotically flat space-times.
Thus we have an infinite chain of universes connected by time-like wormholes,
whose Penrose diagram is shown in Fig.~2. We also have a disjoint
asymptotically flat region containing a naked singularity.

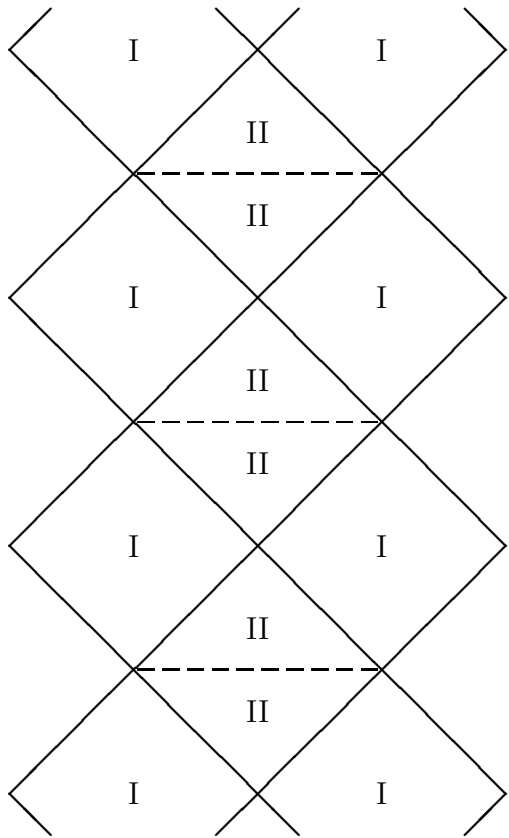
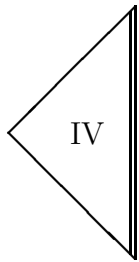
\figtwo{0}{Fig.~2. (a) Penrose diagram for the maximally extended
exact black hole geometry. This consists of an infinite sequence of
asymptotically flat regions I linked by wormholes II. The dashed lines
represent $x=-1$ surfaces. (b) There is also a disjoint asymptotically
flat region IV which contains a naked singularity.}

Such an unusual situation of a black hole space-time having an event horizon
but no singularity has no analog in classical general relativity. But a similar
non-singular geometry has been found in extremally charged black strings in
three dimensions \cite{Horowitz}, which is also an exact string background.
One may be tempted to speculate that non-singular black holes are generic to
string theory, but more examples of such conformally exact metric solutions
of string theory are needed to confirm this.

A direct generalization of the two-dimensional space-time of this letter
to three dimensions (based on the coset space SO(2,2)/SO(2,1)) and four
dimensions (SO(3,2)/SO(3,1)) has also been carried out in Ref.~\cite{BarsI},
and explicit metrics found. However, the complexity of these metrics make
them rather  intractable at this stage. In the $k\rightarrow\infty$ limit,
the three-dimensional metric was found to have a bizarre singularity
structure \cite{BarsII}. Would this simplify in the more realistic case of
finite $k$? Of course other coset space constructions may be considered, and
a classification of possible G/H coset spaces that yield non-compact target
spaces with exactly one time-direction has been done in Ref.~\cite{Ginsparg}.

Ultimately, one would be interested in four-dimensional black holes,
especially representing known solutions of general relativity within a
string framework, and looking for any qualitative differences near the
singularities. However to avoid the presence of singularities, the background
fields of string theory have to violate the strong energy condition, which
is indeed the case for the dilaton.

We intend to relate further details elsewhere.

\bigbreak\bigskip\bigskip\centerline{{\bf Acknowledgement}}
\nobreak\noindent
We wish to thank Stephen Hawking for an interesting conversation.

\bigskip\bigskip

{\renewcommand{\Large}{\normalsize}
}

\end{document}